\documentclass[10pt,conference]{IEEEtran}
\IEEEoverridecommandlockouts
\usepackage{cite}
\usepackage{amsmath,amssymb,amsfonts}
\usepackage{algorithmic}
\usepackage{graphicx}
\usepackage{subcaption}
\usepackage{textcomp}
\usepackage{xcolor}
\usepackage{hyperref}
\usepackage[most]{tcolorbox}
\usepackage{xcolor}

\tcbset{
  rqbox/.style={
    enhanced,
    breakable,
    colback=black!3,     % very light gray background (print-safe)
    colframe=black!60,   % medium gray border
    coltitle=black,
    fonttitle=\bfseries,
    boxrule=0.5pt,       % thinner rule for two-column pages
    arc=1mm,             % subtle rounding (or 0mm for square)
    left=2mm,
    right=2mm,
    top=1.2mm,
    bottom=1.2mm,
    before skip=6pt,
    after skip=6pt,
    attach boxed title to top left={yshift*=-1mm},
    boxed title style={
      colback=black!5,
      colframe=black!60,
      boxrule=0.5pt,
      sharp corners,
      interior style={fill opacity=1},
      left=2mm, right=2mm, top=0.6mm, bottom=0.6mm,
    },
    title={},
  }
}

\def\BibTeX{{\rm B\kern-.05em{\sc i\kern-.025em b}\kern-.08em
    T\kern-.1667em\lower.7ex\hbox{E}\kern-.125emX}}
\begin{document}

\title{What Drives Issue Resolution Speed? An Empirical Study of Scientific Workflow Systems on GitHub\\
}

\author{\IEEEauthorblockN{Khairul Alam}
\IEEEauthorblockA{\textit{Department of Computer Science} \\
\textit{University of Saskatchewan}\\
Saskatoon, Canada \\
kha060@usask.ca}
\and
\IEEEauthorblockN{Banani Roy}
\IEEEauthorblockA{\textit{Department of Computer Science} \\
\textit{University of Saskatchewan}\\
Saskatoon, Canada \\
banani.roy@usask.ca}
}

\maketitle

\begin{abstract}
Scientific Workflow Systems (SWSs) play a vital role in enabling reproducible, scalable, and automated scientific analysis. Like other open-source software, these systems depend on active maintenance and community engagement to remain reliable and sustainable. However, despite the importance of timely issue resolution for software quality and community trust, little is known about what drives issue resolution speed within SWSs. This paper presents an empirical study of issue management and resolution across a collection of GitHub-hosted SWS projects. We analyze 21,116 issues to investigate how project characteristics, issue metadata, and contributor interactions affect time-to-close. Specifically, we address two research questions: (1) how issues are managed and addressed in SWSs, and (2) how issue and contributor features relate to issue resolution speed. We find that 68.91\% of issues are closed, with half of them resolved within 18.09 days. Our results show that although SWS projects follow structured issue management practices, the issue resolution speed varies considerably across systems. Factors such as labeling and assigning issues are associated with faster issue resolution. Based on our findings, we make recommendations for developers to better manage SWS repository issues and improve their quality.

\end{abstract}


\section{Introduction}
In today's research ecosystem, scientific workflows have become a cornerstone for enabling reproducible, scalable, and automated data-intensive experiments \cite{pouchard2019computational}. Scientific workflow systems (SWSs) provide the infrastructure to design, execute, monitor, and share workflows composed of inter-dependent tasks, data transformations, and computational resources \cite{atkinson2017scientific, barker2007scientific}. Examples include platforms such as Galaxy \cite{galaxy2022galaxy}, Nextflow \cite{di2017nextflow}, Pegasus \cite{deelman2015pegasus}, and Snakemake \cite{koster2012snakemake}, which support domains from bioinformatics to remote sensing. These systems allow researchers to model complex analytic pipelines as workflows, manage data dependencies and provenance, automate execution on heterogeneous computing resources (e.g., HPC clusters, cloud), and thus shift attention from manual orchestration toward higher-level scientific insight \cite{liew2016scientific}. However, like any complex software system, SWSs require continuous maintenance. Developers need to fix bugs, implement new features, assist users, update dependencies, and adapt workflows as technologies evolve \cite{alam2023reusability}. Recent studies also highlight that maintaining SWSs presents several challenges, including managing complex dependencies, data structures, and execution environments, handling errors, and ensuring reproducibility \cite{alam2025empirical, alam2025prompt}.

Given these demands, one important aspect of SWS quality and sustainability is how efficiently issues reported by users or contributors are managed and resolved. In open‐source ecosystems such as GitHub, issue trackers provide a mechanism through which bug reports, enhancement requests, questions, and other maintenance tasks are communicated, triaged, addressed, and closed. Prior studies on general software projects have shown that factors such as contributor activity, workload distribution, labeling practices, and the use of structured templates can significantly affect how quickly and effectively issues are resolved \cite{jarczyk2014github, yang2023users, bijlsma2012faster, kuramoto2024understanding}. Faster issue resolution is often associated with higher user satisfaction, stronger community trust, and greater project adoption.

However, SWS projects differ from conventional software in several ways: SWSs are dataflow-oriented platforms that automatically orchestrate computational research pipelines, fundamentally differing from traditional software in their execution model and user focus. They explicitly represent data dependencies as directed acyclic graphs, enabling automatic parallelization across distributed environments, while traditional software uses imperative programming with obscured data flow. Scientific workflows prioritize reproducibility through automatic provenance tracking \cite{alam2022challenges} and are built for petascale data and long-running jobs across HPC and cloud infrastructure. They target domain scientists via visual interfaces and reusable containerized tasks, whereas traditional software emphasizes business logic and developer-centric design. They are often interdisciplinary, serve research-driven purposes, and must manage complex aspects such as data dependencies, execution environments, and workflow reproducibility \cite{alam2025empirical, bharathi2008characterization}. These distinctive characteristics may shape how issues emerge, are prioritized, and are ultimately resolved. Yet, systematic evidence on what drives issue resolution speed in SWSs remains limited. Understanding these factors is crucial for improving the responsiveness, reliability, and long-term sustainability of SWSs.

 In this paper, we seek to fill this gap by conducting an empirical study of issue resolution processes within SWSs hosted on GitHub. Specifically, we aim to understand (1) \emph{how issues are managed and addressed in SWSs}, and (2) \emph{how issue- and contributor-related features relate to issue resolution speed}. To guide this investigation, we formulate the following research questions:
 \begin{itemize}
       \item \textbf{RQ1: How are the issues managed and addressed in scientific workflow systems?} This RQ investigates the overall issue management process, aiming to characterize the practices and strategies used by maintainers and contributors in SWSs.
     \item \textbf{RQ2: What are the relationships between different features and the closure of issues in scientific workflow systems?} This RQ examines how diverse issue- and contributor-related features influence whether and how quickly issues are resolved in SWSs.
 \end{itemize}

Our study analyzes 21,116 issues from 197 repositories related to popular SWSs (e.g., Galaxy, Nextflow) to investigate what drives issue resolution speed on GitHub. Overall, 68.91\% of issues are closed, with half resolved within 18.09 days, though a long tail of unresolved cases indicates persistent maintenance backlogs. Repositories with more contributors do not necessarily resolve issues faster, while those with more total issues tend to take longer to close them. Nextflow shows the best closure rate (86.27\%) and shortest average resolution time (136.42 days), likely due to its centralized governance, whereas Galaxy's distributed structure led to slower responses. Labeling and assignment practices are key management factors; repositories that use them tend to close issues more efficiently. Statistical analysis confirms that while labeling and assignment have small but significant effects on issue closure, textual attributes and contributor counts have negligible influence. These findings highlight that effective triage, coordination, and governance are the main drivers of timely issue resolution in SWSs.

\section{Background and Related Work}
\subsection{Scientific Workflow System (SWS)}
A SWS is an efficient software framework designed to define, execute, and manage complex workflows and associated datasets across diverse computing environments \cite{liu2015survey}. Several well-established SWSs, such as Galaxy \cite{galaxy2022galaxy}, Pegasus \cite{deelman2005pegasus}, Kepler \cite{altintas2004kepler}, Taverna \cite{oinn2004taverna}, Snakemake \cite{koster2012snakemake}, and Nextflow \cite{di2017nextflow}, are now widely adopted across diverse research domains, including bioinformatics and computational biology, medical and health informatics, data science and machine learning, and computational engineering.

\subsection{Issue Resolution in Open Source Software (OSS)}
Prior research \cite{bijlsma2012faster, pfahl2016improving} has shown that issue resolution speed is a critical indicator of project health and developer responsiveness. Studies on OSS repositories have examined how factors such as issue type, reporter expertise, assignee workload, project popularity, and communication activity affect resolution times \cite{bettenburg2008makes, yu2015wait, nguyen2011impact, mockus2002two}. Other works highlight that social and organizational dynamics, like contributor turnover, governance structure, and project maturity, also shape maintenance outcomes \cite{crowston2008free}. Several studies have leveraged GitHub data (e.g., issues, pull requests, commits) to analyze collaboration patterns and productivity. For instance, Gousios et al. \cite{gousios2014exploratory} and Tsay et al. \cite{tsay2014influence} studied pull request acceptance and developer interaction, while Zhou et al. \cite{zhou2016inflow} investigated temporal patterns in issue handling. These studies emphasize that issue resolution is a multifaceted process influenced by both technical (e.g., code complexity, dependencies) and social (e.g., communication, community size) factors. However, despite the growing importance, SWSs remain largely unexplored in this context.

\subsection{Research Gap}
Compared to general-purpose OSS, SWS projects exhibit distinct characteristics, more specialized developer communities, domain-specific user bases, and a strong emphasis on reproducibility and validation \cite{hannay2009scientists, katz2021taking}. SWSs, in particular, must balance usability for domain scientists with technical robustness, which often introduces unique challenges in software maintenance and evolution \cite{alam2025empirical}. However, there is a lack of empirical studies investigating issue management practices and developer responsiveness within this class of projects. This study addresses this gap by analyzing issue-resolution speed across SWS projects and identifying the factors that drive faster resolution.

\section{Study Design}
To examine issue resolution dynamics within SWSs, we focus on three widely adopted platforms: Galaxy\footnote{https://github.com/galaxyproject}, Nextflow\footnote{https://github.com/nextflow-io}, and Snakemake\footnote{https://github.com/snakemake}. A large and active developer community supports these systems and exhibits extensive issue-tracking activity on GitHub. They also embody distinct paradigms of workflow design and execution: Galaxy provides a graphical, web-based environment that simplifies workflow construction for domain scientists \cite{galaxy2022galaxy}; Nextflow introduces a domain-specific language for portable and scalable workflow execution across heterogeneous environments \cite{di2017nextflow}; and Snakemake offers a Python-based framework emphasizing automation and reproducibility \cite{koster2012snakemake}. The diversity of these systems enables a comprehensive examination of issue management practices across varying design philosophies and user communities. Moreover, their open-source nature and long development histories (spanning over a decade) provide transparent, longitudinal datasets suitable for our analysis.

At the time of data collection (September 2025), the Galaxy project supported 147 repositories, Nextflow had 106 repositories, and Snakemake had 63 repositories, resulting in a total of 316 repositories. Among them, 197 contain issues. To answer our RQs, we collect issue data from these 197 repositories using the GitHub API. In total, we retrieved 21,116 issues. For each issue, we capture all available fields, including the issue's title, body, creation and closing timestamps, state (open or closed), labels, and discussion comments. In addition to issue-level information, we collect associated repository metadata, such as the number of stars, forks, and contributors, to characterize project popularity and community engagement.

For RQ1, we aim to understand how maintainers and contributors handle issues. This includes examining issue types, labeling and assignment practices, discussion activity, and closure trends across repositories. The goal is to characterize maintenance workflows and identify patterns in issue management and collaboration practices. For RQ2, we investigate how issue- and contributor-level attributes, such as the presence of labels, assignees, and discussions, relate to whether and how quickly issues are resolved. We conduct statistical comparisons between open and closed issues and use correlation-based modeling to identify features associated with faster resolution. The detailed analytical procedures and results for each RQ are presented in the section \ref{results}.

\smallskip
\noindent The \textbf{replication package} is available in online appendix \cite{random_2025_17555352}.
\section{Results}
\label{results}
\subsection{RQ1: How are issues addressed and managed?}
\begin{figure}[htbp]
    \centering
    \begin{subfigure}[b]{0.32\columnwidth}
        \centering
        \includegraphics[width=\textwidth]{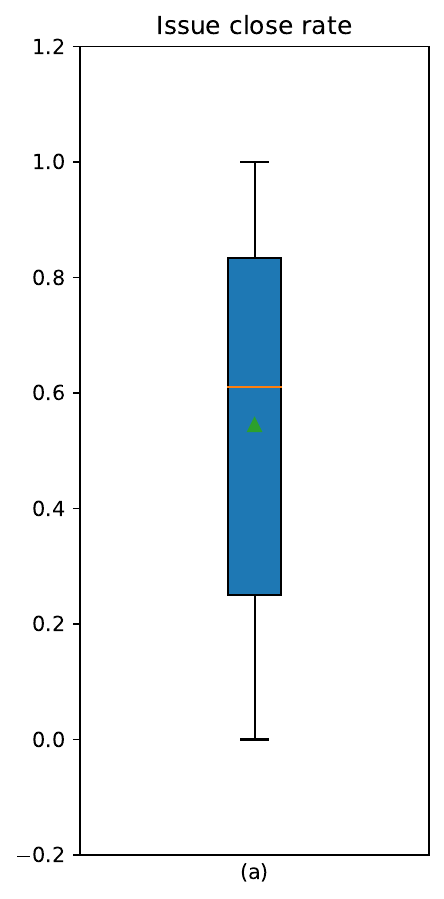} 
        \label{fig:1a}
    \end{subfigure}
    \hfill
    \begin{subfigure}[b]{0.32\columnwidth}
        \centering
        \includegraphics[width=\textwidth]{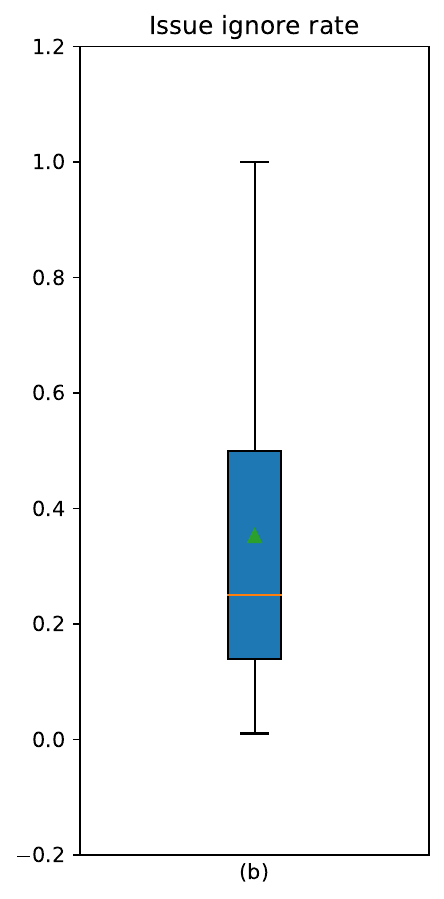} 
        \label{fig:1b}
    \end{subfigure}
    \hfill
    \begin{subfigure}[b]{0.32\columnwidth}
        \centering
        \includegraphics[width=\textwidth, height=0.243\textheight]{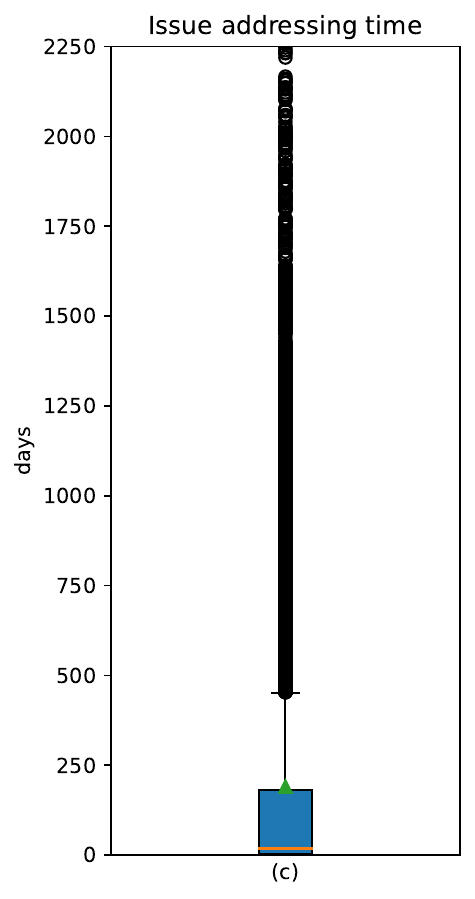}
        \label{fig:1c}
    \end{subfigure}
  \caption{The distribution of closed issues rate (a), ignored issues rate (b), and addressing time (c) in investigated repositories.}
  \label{fig:issue-rate-comparison}
\end{figure}

\textbf{Motivation: }Effective issue management is essential for the quality, usability, and sustainability of SWSs. On GitHub, issues function as the primary medium for reporting bugs, requesting features, and coordinating maintenance, reflecting both project maturity and community health. SWSs involve interdisciplinary contributors with varying expertise, leading to diverse issue types such as bug fixes, dependency management, documentation updates, system redesign, and API migration \cite{alam2025empirical}. While some issues are easily resolved, others require extensive coordination and long-term planning, and neglecting them can hinder the scientific use of SWSs. As the SWS ecosystem continues to grow, understanding issue management practices is critical for improving maintainability and fostering sustainable communities.
\newline
\textbf{How are the issues addressed?} We aim to examine how issues are managed and addressed in SWSs. These issues are divided into open and closed categories. Using the GitHub API, we collect a total of 21,166 issues from the selected repositories. Of these, 6,581 (31.09\%) are open, and 14,585 (68.91\%) are closed. On average, each repository contains 107.44 issues. Notably, 36 repositories (18.27\%) have closed all their issues. Furthermore, 44 repositories (22.34\%) exhibit a closed issue rate (\textbf{CIR}) exceeding 90\%, with an average of 17.25 issues each. This indicates that repositories with a \textbf{CIR} greater than 0.9 generally have a smaller number of total issues compared to others. The results of the closed issue rate are presented in Figure \ref{fig:issue-rate-comparison}(a).

We find that 25.40\% of closed issues are \emph{self-closed}, meaning that the individuals who raised them can resolve them, often through subsequent discussion. In addition, we identify a category of issues referred to as \emph{ignored issues}, those that receive no response or engagement from other users or developers. These ignored issues account for 11.23\% (2,378) of all issues and 36.13\% of open issues. We deﬁne the \emph{ignored issue
rate} as the ratio of ignored issues for each repository. This metric provides insight into the level of maintainer and community engagement in addressing issues. The distribution of the ignored issue rate is shown in Figure \ref{fig:issue-rate-comparison}(b).

Additionally, we examine the addressing time of issues, defined as the duration between an issue's creation and its closure. Across the 14,585 closed issues, the average addressing time is 190.65 days with a standard deviation of 397.11 days. However, the median addressing time is 18.09 days, suggesting that while most issues are resolved relatively quickly, a few long-standing cases extend the overall average. Notably, the longest addressing time exceeds four years. The distribution of addressing times is shown in Figure \ref{fig:issue-rate-comparison}(c). 

To investigate the factors influencing issue resolution time, we analyze two repository-level characteristics: (1) the number of contributors and (2) the total number of issues. While repositories with more contributors might be expected to resolve issues faster due to a larger available workforce, our analysis reveals otherwise. The number of contributors exhibits a statistically significant (\( p < 0.01 \)) but negligible effect, suggesting that contributor count does not meaningfully improve issue resolution speed. In contrast, repositories with a greater number of issues tend to have longer addressing times, as increased workload can delay resolution. This observation is supported by a weak yet statistically significant positive correlation between the total number of issues and average addressing time  (\( p < 0.01 \)).
\newline
\textbf{How are the issues managed?} GitHub provides several mechanisms to help maintainers manage issues. In this study, we focus on two primary methods: (1) using labels to tag issues and (2) assigning issues to specific contributors for resolution. Throughout this section, we refer to these practices as \emph{labeling} and \emph{assigning}, respectively. 
\begin{figure}[htbp]
  \centering
    \includegraphics[width=\linewidth]{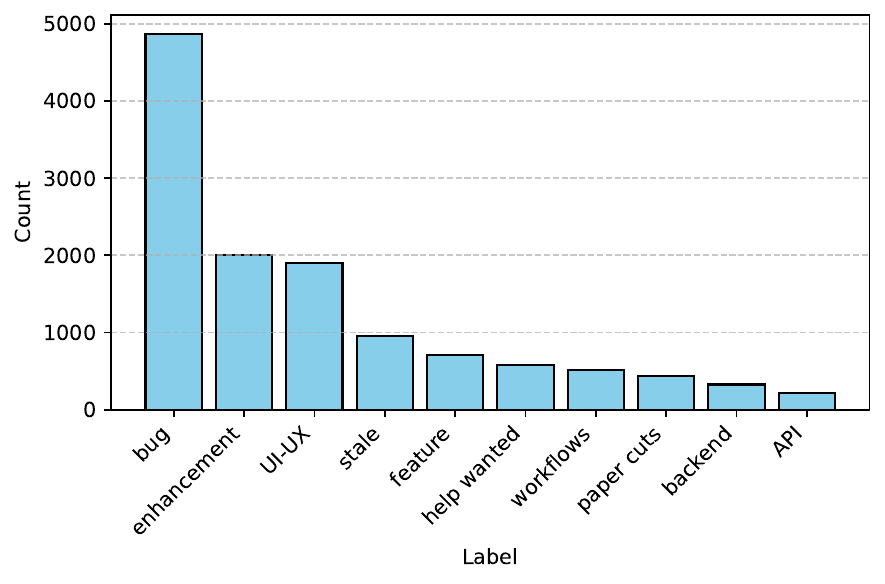}
    \vspace{-1.5em}
  \caption{The distribution of top 10 most frequently used labels.}
  \label{fig:labels-rank}
\end{figure}

Our analysis shows that 40.61\% (80) of the repositories utilize labels to organize issues, while 50.76\% (100) assign issues to assignees. At the issue level, 54.09\% (11,449) of issues include one or more labels, and 20.39\% (4315) have assigned contributors.
GitHub provides a standard set of nine default labels, such as \emph{bug, duplicate, enhancement, help wanted, invalid, question, wontfix, and good first issue}. However, active issue management often involves defining custom labels beyond these defaults. Across all analyzed repositories, we identify a total of 296 unique labels (including the default ones). We then calculate the frequency of each label to examine labeling trends. Figure \ref{fig:labels-rank} presents a ranked distribution of the most commonly used labels.

\begin{tcolorbox}[rqbox]
In our dataset, 68.91\% of issues are closed, with a median resolution time of 18.09 days. At the repository level, 40.61\% of repositories use labels to categorize issues, and 50.76\% assign issues to contributors. At the issue level, 54.09\% of issues carry one or more labels, while 20.39\% have at least one assignee. Notably, among the top ten most frequently used labels, only three correspond to GitHub's default set, while seven are community-defined (e.g., workflows, paper cuts, backend)
\end{tcolorbox}

\subsection{RQ2:Relationships between different features and the closure of issues}
\textbf{Motivation: }This RQ aims to explore the relationships between various features of issues and their closure in SWS repositories. SWSs are developed within interdisciplinary teams that include domain scientists and software engineers, leading to diverse development practices and maintenance challenges. Understanding how factors such as issue characteristics, contributor activity, and project attributes relate to issue closure can shed light on what drives efficient problem resolution in these projects. The answer to this RQ can provide practical suggestions for both repository maintainers and issue raisers to improve the management, prioritization, and resolution of issues in SWS repositories.

\textbf{Investigated Features. }We examine the relationship between seven key features and the likelihood of issue closure. The first six features describe issue-level characteristics, while the seventh feature, \textit{num\_contributors}, captures a repository-level attribute. The seven features considered in our analysis are presented as follows:
\begin{itemize}
    \item \emph{has-label: }Indicates whether an issue has been labeled. On GitHub, users with triage access can tag issues with labels. A labeled issue typically suggests that it has been reviewed or acknowledged by maintainers, potentially increasing its likelihood of being addressed.
    \item \emph{has-assignee: }Denotes whether an issue has been assigned to a specific contributor. Users with write permissions can assign issues to developers, making the assignee responsible for resolving the issue. Assigned issues are therefore more likely to reach closure.
    \item \emph{title-length: }Measures the length of the issue title. An
informative title may help the repository maintainers
better understand the issue at a glance. Usually, a longer title carries more information that can catch the maintainer's attention and facilitate faster responses.
    \item \emph{body-length: }Refers to the length of the issue description. A detailed and well-written description provides maintainers with essential context and clarity, increasing the chances of timely resolution. Many repository guidelines explicitly encourage such comprehensive reporting.
    \item \emph{has-code: }Identifies whether the issue body includes code blocks. Code snippets, such as reproducing examples or error traces can help maintainers understand and reproduce the reported problem, aiding resolution.
    \item \emph{has-url: }Determines whether the issue description contains URLs. Using Markdown syntax, reporters may link to related issues, web resources, or images. These references can supply valuable information for maintainers.
   \item \emph{num-contributors: }Represents the total number of contributors in a repository; repositories with more active contributors are generally better capable of addressing issues efficiently.
\end{itemize}

\textbf{Feature Measurement and Statistical Test. }We measure the length of each issue title and body by applying the built-in Python \emph{len()} function. Using regular expressions, we further detect whether the issue description contained code blocks or URLs. For each issue, we extract the six structural and management-related features, including \emph{has-label}, \emph{has-assignee}, \emph{title-length}, and \emph{body-length}. The number of contributors in a repository is obtained by querying the GitHub API to fetch metadata about the repository. Finally, to determine whether these feature distributions significantly differed between open and closed issues, we apply the non-parametric \emph{Wilcoxon rank-sum test} \cite{wilcoxon1945individual}.

\begin{table}[htbp]
\footnotesize
\centering
\caption{The significance of the differences between the feature values of open and closed issues}
\label{tab:feature_significance}
\begin{tabular}{lcc}
\hline
\textbf{Feature} & \textbf{P-value} & \textbf{Cohen's $\delta$} \\
\hline
has-label        & $<$0.01 & 0.2 (Small) \\
has-assignee     & $<$0.01 & 0.23 (Small) \\
title-length     & $<$0.01 & 0.07 (Negligible) \\
body-length      & $<$0.01 & 0.08 (Negligible) \\
has-code         & $<$0.01 & 0.12 (Negligible) \\
has-URLs         & $<$0.01 & 0.02 (Negligible) \\
num-contributors & $<$0.01 & 0.06 (Negligible) \\
\hline
\end{tabular}
\end{table}

In addition to statistical significance testing, we also compute Cohen's effect size ($d$) to quantify the magnitude of differences between open and closed issues. While \emph{p-values} indicate whether the observed differences are statistically significant, effect sizes capture their practical impact, that is, how meaningful these differences are in real-world terms. Following established conventions, values of $|d| < 0.2$ are considered negligible, $0.2 \leq |d| < 0.5$ as small, $0.5 \leq |d| < 0.8$ as medium, and $|d| \geq 0.8$ as large.

 Table~\ref{tab:feature_significance} summarizes the statistical comparison between open and closed issues across all investigated features. All features show statistically significant differences (\( p < 0.01 \)), indicating that each characteristic varies meaningfully between open and closed issues. However, the effect size values ($d$) reveal that the magnitude of these differences is generally small or negligible, suggesting that while the differences are statistically detectable, their practical impact is limited.

However, the presence of labels and assigned developers (i.e., \emph{has-label and has-assignee}) show small but meaningful effects ($d$ = 0.20 and 0.23, respectively). This finding aligns with expectations that labeled and assigned issues are more likely to attract attention and thus reach closure. These two features appear to play a modest but tangible role in issue management effectiveness.
In contrast, textual and contextual attributes, such as \emph{title-length, body-length, has-code, and has-URLs} exhibit negligible effect sizes ($d$ $<$ 0.2), implying that the amount of written content or inclusion of supporting information (e.g., code snippets, URLs) does not substantially differentiate open from closed issues. Similarly, the number of contributors in a repository shows only a negligible effect, indicating that having more contributors does not necessarily lead to faster or more consistent issue resolution.

Overall, these results suggest that organizational and management-related features (e.g., labeling and assignment) are more influential for issue closure than content-related or repository-level features, which have minimal practical impact despite statistical significance.

\begin{tcolorbox}[rqbox]
Our analysis shows that management-related factors, particularly labeling and assignment, have the strongest association with issue closure in SWS repositories, albeit with small effect sizes. In contrast, content-related attributes (e.g., title length, body length, code snippets, URLs) and repository-level characteristics (e.g., number of contributors) show negligible influence.
\end{tcolorbox}

\section{Discussion \& Implications}
Our analysis shows that SWSs exhibit mature yet uneven issue management practices. Although the median time-to-close of 18.09 days reflects responsive maintenance, a long tail of unresolved or delayed issues, some persisting for years, reveals substantial maintenance debt. 68.91\% of issues are closed, but the persistence of backlogs indicates ongoing challenges in workload distribution and prioritization.
\begin{table}[htbp]
\centering
\caption{Comparison of SWS Repositories and Issues}
\label{tab:sws_comparison}
\footnotesize
\begin{tabular}{lccc}
\hline
\textbf{Metric} & \textbf{Galaxy} & \textbf{Nextflow} & \textbf{Snakemake} \\
\hline
No. of Repos & 147 & 106 & 63 \\
Repos With Issues & 104 & 48 & 45 \\
Total Issues & 13{,}972 & 4{,}509 & 2{,}685 \\
Avg. Issues/Repo & 134.35 & 93.94 & 59.67 \\
Avg. Stars & 36.58 & 104.38 & 74.44 \\
Avg. Forks & 44.71 & 37.94 & 24.98 \\
Avg. Watchers & 15.09 & 5.79 & 2.00 \\
Avg. Contributors & 14.50 & 10.44 & 8.82 \\
Avg. Close Time (days) & 215.65 & 136.42 & 175.34 \\
Close Rate (\%) & 66.53 & 86.27 & 52.10 \\
Assign Rate (\%) & 26.17 & 8.58 & 10.09 \\
Label Rate (\%) & 50.34 & 52.63 & 76.05 \\
\hline
\end{tabular}
\end{table}
Cross-system comparisons (Table \ref{tab:sws_comparison}) highlight several differences. Nextflow demonstrates the fastest average closure time (136.42 days) and highest close rate (86.27\%). In contrast, Galaxy's distributed ecosystem, spanning over a hundred repositories maintained by heterogeneous contributors, leads to slower resolution (215.65 days) and a higher ignored-issue rate. Snakemake falls between these extremes. These disparities underscore that project coordination and governance models, rather than repository size or community scale, primarily drive responsiveness within SWS ecosystems.
Feature-level analyses further reveal that management-related factors, particularly labeling and assignment, exert a stronger influence on issue closure than contextual attributes such as title length, body length, or code snippets. The substantial proportion of self-closed and ignored issues also suggests that many reports stem from user misunderstanding or documentation gaps rather than code defects, emphasizing the need for better maintainer engagement and clearer communication practices.

\textbf{Practical implications.}
Maintainers should (1) enforce consistent labeling and early assignment to ensure issue visibility, (2) employ lightweight automation for triage and stale-issue detection, and (3) strengthen user documentation to reduce self-closure and repetition. For platform designers, integrating analytics dashboards that track issue aging can support proactive backlog management. For the broader SWS community, establishing shared maintenance guidelines and contributor training could enhance long-term sustainability.

\textbf{Research implications.}
Future studies can model causal factors of resolution time through regression or survival analysis, and complement quantitative patterns with qualitative insights from maintainers. Extending this work to other research-software ecosystems may clarify whether responsiveness challenges are intrinsic to interdisciplinary, data-intensive projects.

\section{THREATS TO VALIDITY}
\textbf{Internal Validity: }Our analysis is correlational and does not establish causality between issue characteristics and resolution speed. For instance, while labeled issues tend to close faster, confounding factors, such as contributor experience, project governance, or issue complexity, may also influence results. To mitigate such risks, we employ nonparametric statistical tests and effect-size analyses, interpret results conservatively, and emphasize association rather than causation.

\textbf{Construct Validity: }For analytical convenience, we treat closed issues as resolved. However, this assumption may not always hold, as issues can be closed without being fully addressed (e.g., duplicates, out of scope), which may introduce some measurement inaccuracy.

\textbf{External Validity: }It addresses the generalizability of our findings. We focus on three mostly used SWSs, which represent diverse paradigms of workflow design. However, results may not generalize to other SWSs, particularly smaller or less active ones. Future studies can extend our analysis to additional SWSs to enhance generalizability.

\section{Conclusion}
This study examines what drives issue resolution speed in SWSs hosted on GitHub. By analyzing 21,116 issues across prominent SWSs, Galaxy, Nextflow, and Snakemake, we provide an empirical view of how issue management practices influence responsiveness and maintenance efficiency. Our results show that while SWS projects generally exhibit structured and mature issue-handling processes, resolution speed varies considerably across systems. We find that management-related practices, particularly labeling and assigning issues, are significantly associated with faster closure, whereas content-related attributes (e.g., description length or inclusion of code snippets) and the number of contributors have minimal impact. These findings emphasize the importance of effective triage, task assignment, and governance structures in sustaining healthy open-source scientific software communities.

In the future, we plan to investigate how project-level characteristics, such as maturity, activity level, and workload distribution, affect issue resolution speed across different SWSs. Expanding the analysis to include additional SWSs and related research software domains will enhance the generalizability of our findings. Moreover, integrating quantitative modeling with qualitative insights from maintainers can help uncover the causal and organizational factors that drive responsiveness within SWS communities.

\bibliographystyle{ieeetr}
\bibliography{Bibliography}

\end{document}